# Assessing the brittle crust thickness from strike-slip fault segments on Earth, Mars and Icy Moons

Frédéric-Victor Donzé [1,*], Yann Klinger [2], Viviana Bonilla-Sierra [1], Jérôme Duriez [3], Liqing Jiao [2] and Luc Scholtès [4]

[1] Univ. Grenoble Alpes, Univ. Savoie Mont Blanc, CNRS, IRD, IFSTTAR, ISTerre, 38000 Grenoble, France ;
[2] Institut de Physique du Globe de Paris, Sorbonne Paris Cite, Université Paris Diderot, UMR 7154, CNRS, Paris, France ;
[3] Inrae, UR RECOVER, Aix-en-Provence F-13182, France ;
[4] Université de Lorraine, CNRS, GeoRessources, F-54000 Nancy, France;
[*] Correspondence: frederic.donze@univ-grenoble-alpes.fr;

**Abstract:** Segment lengths along major strike-slip faults exhibit a size dependency related to the brittle crust thickness. These segments result in the formation of the localized "P-shear" deformation crossing and connecting the initial Riedels structures (i.e. en-echelon fault structures) which formed during the genesis stage of the fault zone. Mechanical models show that at all scales, the geometrical characteristics of the Riedels exhibit dependency on the thickness of the brittle layer. Combining the results of our mechanical discrete element model with several analogue experiments using sand, clay and gypsum, we have formulated a relationship between the orientation and spacing of Riedels and the thickness of the brittle layer. From this relationship, we derive that for a pure strike-slip mode, the maximum spacing between the Riedels are close to three times the thickness. For a transtensional mode, as the extensive component becomes predominant, the spacing distance at the surface become much smaller than the thickness. Applying this relationship to several well-characterized strike-slip faults on Earth, we show that the predicted brittle thickness is consistent with the seismogenic depth. Supposing the ubiquity of this phenomenon, we extent this relationship to characterize en-echelon structures observed on Mars, in the Memnonia region located West of Tharsis. Assuming that the outer ice shells of Ganymede, Enceladus and Europa, exhibit a brittle behavior, we suggest values of the corresponding apparent brittle thicknesses.

## 1. Introduction

Several major strike slip faults exhibit a succession of rupture induced segments associated on the ground surface, which present a maximum characteristic length as revealed through piecewise linear fitting (Klinger, 2010; Lefevre et al., 2020). Measurements of the maximum horizontal extent of individual slip-patches in strike-slip systems show that their strike length reaches maximum values. For example, this length ranges from 10 to 30 km along the central part of the San Andreas Fault (Wallace, 1973; Scholz, 1998). Recent studies suggest that a relationship might exist between the thickness of the brittle crust and the length of these fault segments (Klinger, 2010; Cambonie et al., 2018). Moreover, fault systems exhibiting regular spatial patterns like the San Andreas and the Walker faulted zones, could suggest that similar relationships might exist at the scale of a fault set (Zuza et al., 2018).

From small-scale experiments, it has been shown that at embryonic stages of a shear fault zone formation, oblique evenly spaced fractures known as en-echelon fractures or Riedel shear structures (also referred to as R shears), appear at the surface (Naylor et al., 1986; Xiao et al., 2017; Atmaoui et al., 2005; Lefevre et al., 2020; Ueta et al., 2000) (Figure 1-a,b,c,d,e).

For a pure strike slip fault, which corresponds mechanically to a pure anti-plane shear deformation, a hierarchical process takes place, controlled by a series of bifurcations (Goldstein and Osipenko, 2012) (Figure 1-f). This means that a dense fracture system is first generated at the tip of the parent basal displacement zone (Gopalakrishnan and Mecholsky, 2014) due to the strong interaction between the cracks emerging from the shear deformation (Du and Aydin, 1991). Then, some of these initial cracks coalesce to produce well-formed fractures, which propagate independently from each other (Simón et al., 2006; Lin et al., 2010). Their growth is unstable at the initial stage and stabilizes afterwards; this is due to the relaxation of the surrounding stress field produced by the propagation of the neighboring fractures (Goldstein and Osipenko, 2012). Ultimately, only some of them continue to grow to eventually form an en-echelon fracture set at the ground surface, exhibiting regular intervals S and with a twist angle ω (Figure 2), which correspond to a multiple of the distance between the numerous fractures at depth (Goldstein and Osipenko, 2012).

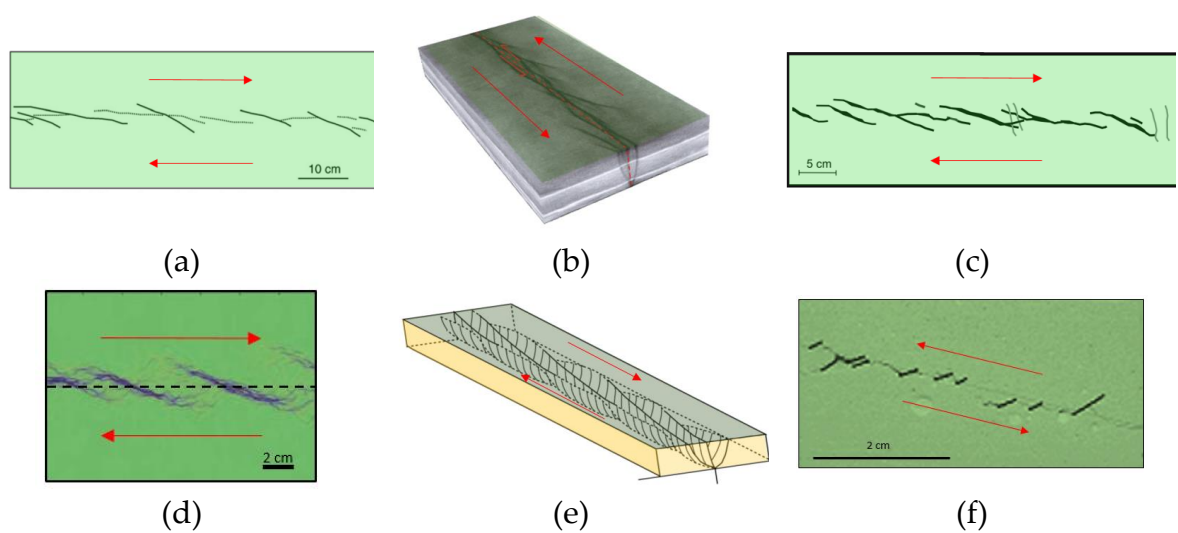

**Figure 1.** Some experimental models of the formation of en-echelon structures: (a) Sand box model (Naylor et al., 1986); (b) Sand box model (Xiao et al., 2017); (c) Clay model (Atmaoui et al., 2005); (d) Sand box model (Lefevre et al., 2020); (e) Sand box model (Ueta et al., 2000); (f) Gypsum model (Goldstein and Osipenko, 2012).

As shearing develops, it has been observed that the initial en-echelon patterns become inactive and the deformation localizes along a narrow zone called P-shear (Naylor, 1986; Atmaoui et al., 2005; Lefevre et al., 2020) (Figure 2-a). This P-shear, can cross or absorb the Riedel structures, forming relay zones (Figure 2-b,c,d,e).

In cases of transtensive deformation, the spacing *S* and the twist angle *ω* of the Riedels (also noted R shears) appearing at the surface can be subjectto major modifications (Pollard et al., 1982; Leblond and Frelat, 2014; Pons and Karma, 2010). Thus, the resulting organization of the en-echelon fractures will not only depend on the thickness of the brittle layer as mentioned previously but also on the mode of deformation.

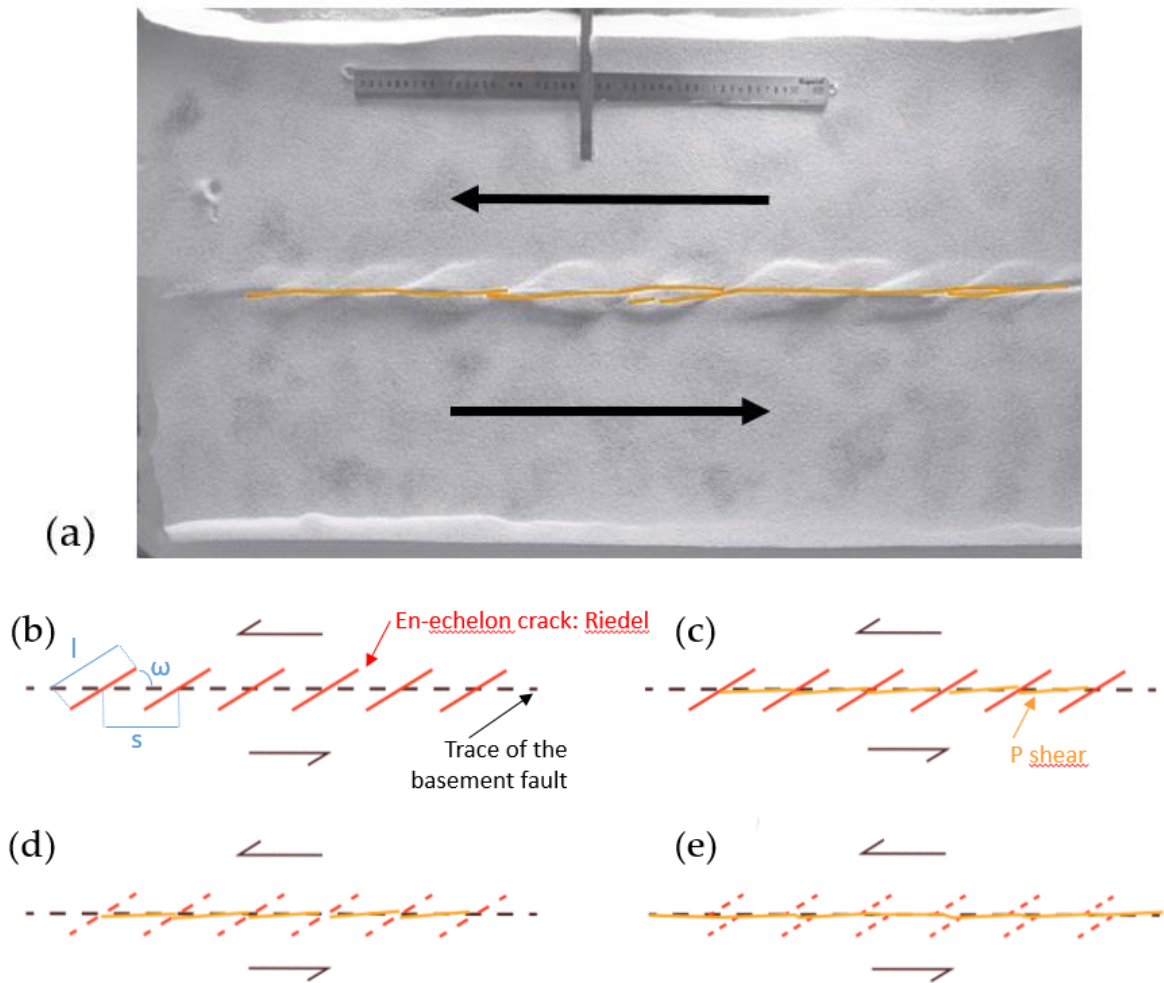

**Figure 2.** (a) Map of the active shears in a sand box experiment (Lefevre, 2020), when the P-shears have already linked to form a continuous fault. (Bottom) Evolutionary scheme of the formation of strike-slip faults, seen from the surface: (b) apparition of the Riedels where l, s and ω correspond to the length, spacing and twist angles of the Riedels respectively; (c) the P-shears reach the surface and take place between successive Riedels; (d) P-shears coalesce and form a continuous fault; (e) the Riedels are no longer active and are passively translated as the continuous fault exhibits segments with a characteristic length related to the initial spacing length (adapted from Lefevre, 2018).

To investigate the relationships between the twist angle ω and the spacing $S$ between the en-echelon fractures, the thickness of the brittle layer and the mode of deformation (transtensive or transpressive modes), we carried out numerical simulations based on a discrete element model implemented in the Yade DEM open source software (Kozicki & Donzé, 2008, 2009). We first used the model to study the evolution of the twist angle of an en-echelon fracture set as the transtensive loading mode, which evolves from pure extensive to pure strike slip modes. We compared our results with two formulations proposed by Pollard et al. (1982) and Leblond and Frelat (2014). We then compared our numerical results with results obtained from previous small scale experiments performed in clay materials (Tchalenko, 1970; Clifton et al., 2000; Schlische et al., 2002), sand materials (Naylor et al., 1986; Zuza, 2017; Lefevre et al., 2020), Walnut experiments (Zuza, 2017) and gypsum (Goldstein and Osipenko, 2012). Based on these results, we derive a formulation, which links the spacing between the Riedels with the thickness, depending on their twist angles. We then apply this formulation to predict the thickness of a crustal brittle layer crossed by an active fault exhibiting en-echelon pattern at surface (i.e., the Greendale fault). We also discuss the predictions of this formulation in light of the corresponding seismogenic depth of several major strike-slip faults for which the fault segments size might correspond to the P-shear (**Figure 2-c**) appearing during their genesis process. Finally, we consider en-echelon systems identified on Mars and Icy moons to assesthe corresponding thickness of the brittle layers through which they propagate.

## 2. Numerical model

To simulate strike-slip faulting within a brittle cohesive-frictional layer, we the bonded particle formulation as proposed by (Scholtes and Donzé, 2013) and implemented in the YADE DEM platform (Kozicki and Donzé, 2008, 2009; Šmilauer et al., 2010). The brittle medium is modeled as an assembly of spherical elements interacting with each other according to predefined elastic-brittle interaction laws. Newton's second law of motion is applied to calculate the acceleration from which the discrete elements velocities and displacement are obtained by successive integrations based on an explicit finite difference scheme. In order to keep the numerical scheme stable, the iterative time step was set small enough to avoid any numerical oscillations and a local non-viscous type damping was used to dissipate kinetic energy to ensure each iterative step was close to a quasi-static equilibrium.

### 2.1. Formulation of the model

In discrete element models (DEM), the global mechanical behavior is controlled by the local constitutive laws defined at the particle contacts. In the classic granular description of the medium, only particles in direct contact can interact. The approach used here is slightly different as an interaction range enables near neighbor interactions between discrete elements, which are not-strictly in contact (Scholtès & Donzé, 2013). A coefficient $\gamma_{int}$ ($\gamma_{int} \geq 1$) characterizes the distance over which particles may interact (**Figure 3-a**). Using this non-local approach, the proposed model is able to simulate macroscopic behaviors specific to elastic-cohesive-fictional brittle rocks characterized by high ratios of compressive to tensile strengths and non-linear failure envelopes.

The interaction forces were computed from the constitutive laws proposed by Scholtès and Donzé (2013). The normal force $F_n$ (Figure 3-b), is given by,

$$F_n = k_n \ \Delta D \qquad (1)$$

where $k_n$ is the normal stiffness calculated as

$$k_n = 2E_{eq} \frac{R_a R_b}{(R_a + R_b)} \qquad (2)$$

Here, $\Delta D$ is the relative displacement between the interacting discrete elements, defined as $\Delta D = D - D_{eq}$, where $D$ is the distance between the centroids of the interacting elements and $D_{eq}$ is the initial equilibrium distance. $E_{eq}$ is the local equivalent of the Young's modulus, and $R_a$ and $R_b$ are the radii of the two particles in contact. In compression, $F_n$ can increase linearly with $\Delta D$. In tension, a maximum acceptable force $F_{n,max}$ is defined as a function of the interaction tensile strength $t$, such that, $F_{n,max} = -tA_{int}$, where $A_{int} = \pi(_{min}(R_a, R_b))^2$ is the interacting surface area between the discrete elements $a$ and $b$. When $F_n$ reaches its maximum value, tensile rupture occurs at the contact location and the interaction is then deleted from the computational scheme.

The shear force $F_s$ is computed incrementally by updating its orientation and intensity according to:

$$F_s = \{F_s\}_{updated} + k_s \Delta u_s \qquad (3)$$

with $\Delta u_s$ the relative incremental tangential displacement and $k_s$ the shear stiffness which are calculated based on the local stiffness ratio $\mu$ and on the normal stiffness $k_n$ such that:

$$k_s = \mu \ k_n \qquad (4)$$

Following a Mohr-Coulomb type criterion, the maximum admissible shear force $F_{s,max}$ depends on the normal force, $F_n$, the cohesion $c$ and the local peak frictional angle $\varphi_b$ and is calculated as:

$$F_{s,max} = F_n tan\varphi_b + cA_{int} \qquad (5)$$

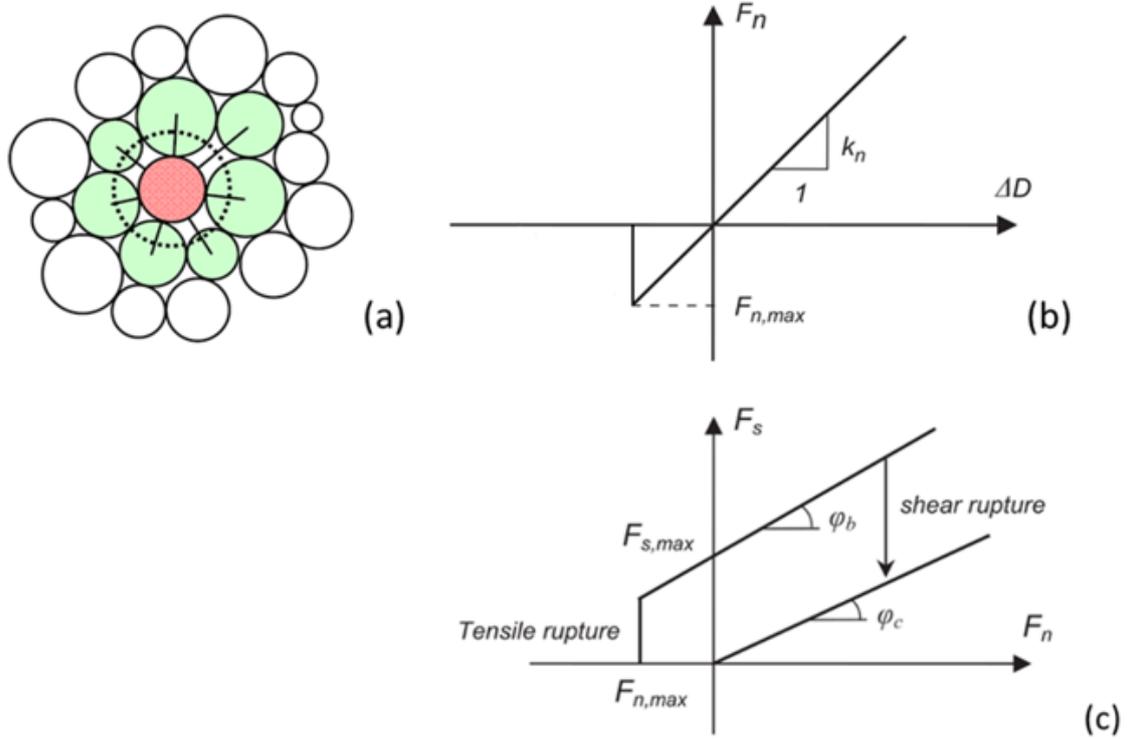

**Figure 3.** (a) Sketch of the effect of interaction distance on the contact fabric with $\gamma_{int} > 1$, (b) Normal interaction force between discrete elements and (c) rupture criterion modified from Scholtès and Donzé (2013).

When $F_s \geq F_{s,max}$, a shear rupture occurs at the contact location and the interaction is then deleted from the computational scheme (Figure 3-c).

If debonded particles come into contact afterwards, the normal force according to Equation (1). with $F_{n,max} = 0$, and the maximum admissible shear force becomes:

$$F_{s,max} = F_n tan\varphi_c \qquad (6)$$

In the present study, $\varphi_b = \varphi_c$. S The physical and mechanical properties used in the model are given in **Table 1**.

Table 1. Model parameters.

| $\gamma_{int}$ | Density (kg/m³) | Gravity | Young's modulus, $E_{eq}$ (GPa) | ks/kn | Friction angles, $\varphi_b = \varphi_c$. | Tensile strength, t (MPa) | Cohesion c (MPa) |
|---|---|---|---|---|---|---|---|
| 1.3 | 2500 | 10 | 15 | 0.3 | 18 | 4.5 | 45 |

*2.2. Model set-up*

To investigate the relationships between the en-echelon fractures' spacing, their orientations and the thickness of the brittle layer, we set up a strike-slip experiment based on a parallelepipedic shear box entirely composed of discrete elements (**Figure 4-a**) The parallelepiped box has a length *L*, a width *W* = 0.89x*L* and a reference thickness T*ref* = *0.083*x*L*. The deformation is controlled by the displacements imposed on the discrete elements composing the basal plane and the vertical walls. For a pure strike slip configuration, the discrete elements belonging to the blue boundary set move in the direction of the y horizontal axis (blue arrow direction), whereas the ones belonging to the red boundary set move in the opposite direction, i.e. the red arrow direction. A notch located at the basement of the sample and oriented along the deformation axis y, represents the basal parent fault **Figure (4-b)**. The presence of the notch does not modifiy the resulting en-echelon pattern visible at the surface but it decreases the extra amount of cracks occurring near the basement in the transpressive configuration, which masks the shape of the fractures.

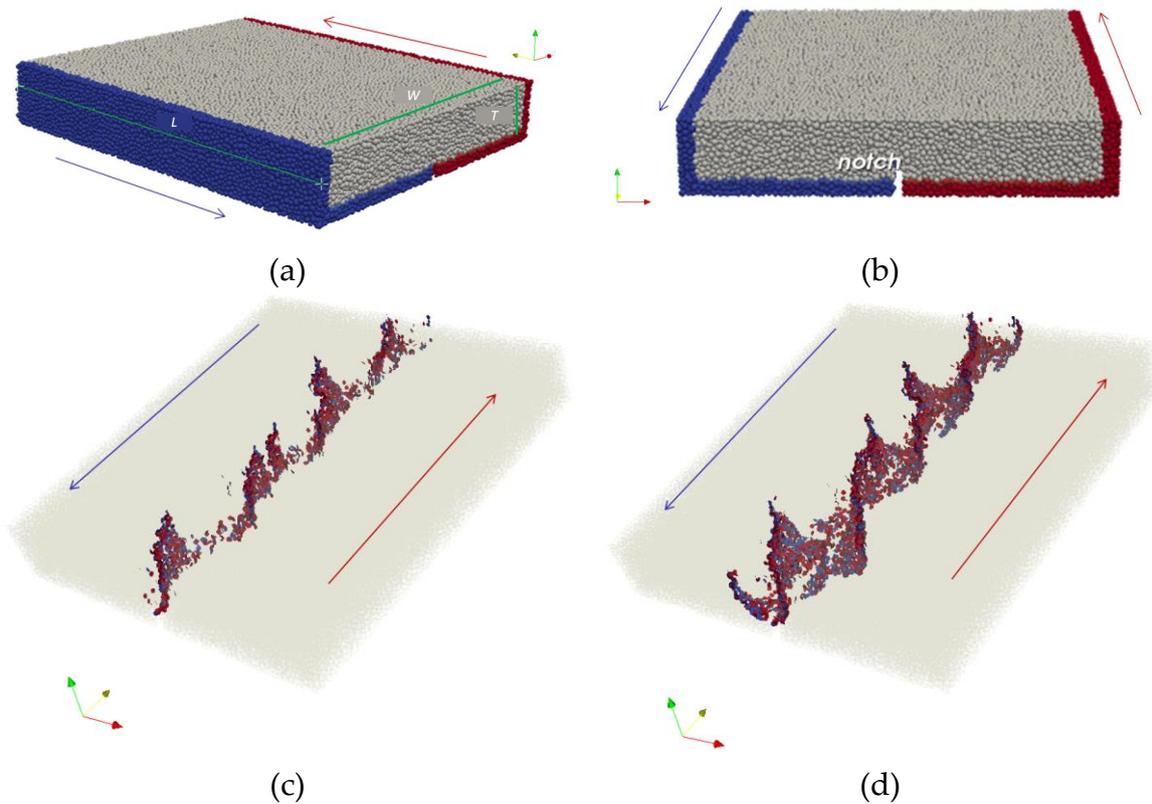

**Figure 4**. Set up of the numerical model, used to simulate the propagation of en-echelon fractures from a basal parent fault in an elastic brittle layer with x,y,z axes are in red, yellow and green color respectively; (**a**) The perspective view shows the boundary conditions imposed to the model. Colored discrete elements are subjected to an imposed displacement along the fault direction to simulate the shearing process: red discrete elements move in the opposite direction to the blue ones (along the yellow axis y); (**b**) Face view showing the presence of the notch at the basement; (**c,d**) The perspective views show the evolution of the fractures' propagation. Red and blue dots represent local cracks due to mode I and mode II failures respectively, at two different deformation stages.

As the shear deformation process progresses, cracks nucleate at the upper tip of the notch and coalesce to form fractures of helical shape (**Figure 4**-**c**). They then self-organize to form an array of en-echelon fractures at the surface of the shear box (**Figure 4**-**d**) exhibiting the characteristic twist angle at the surface (**Figure 2-b)**. As predicted by hierarchical models, some of these fractures prevent others from propagating, so that only some of them actually reach the surface of the model. Most local cracks occur in opening mode I (red dots in **Figure 4-c,d**) and only a few of them occur in shear mode II (blue dots), as expected for the formation of dilatant en-echelon fractures (Pollard et al., 1982).

In order to identify the first order parameters controlling the geometry of the en-echelon patterns at the surface, we carried out simulations considering (i) different thicknesses of the brittle layer for pure strike-slip mode and (ii) different trajectories of the discrete elements forming the driven boundaries in order to produce mixed-mode deformation, i.e. transtensional and transpressional modes.

2.3 Numerical results

2.3.1 Influence of the thickness T

To investigate the influence of the thickness T on the en-echelon pattern for a Pure Strike Slip (PSS) loading, we simulated layers with different thicknesses: a reference thickness $T_{ref}$ (PSS_$T_{ref}$), 0.66 times $T_{ref}$ (PSS_0.66$T_{ref}$), 1.33 times $T_{ref}$ (PSS_1.33$T_{ref}$) and 2 times $T_{ref}$ (PSS_2$T_{ref}$). The results of all those tests are presented in **Figure 5**, where all the induced cracks are colored in black. The images correspond to top views in transparency, i.e. all cracks inside the model are visible. Since the en-echelon fractures rotate near the parent crack tip and propagate vertically, the dark aligned narrow zones correspond to the en-echelon fractures visible at the surface.

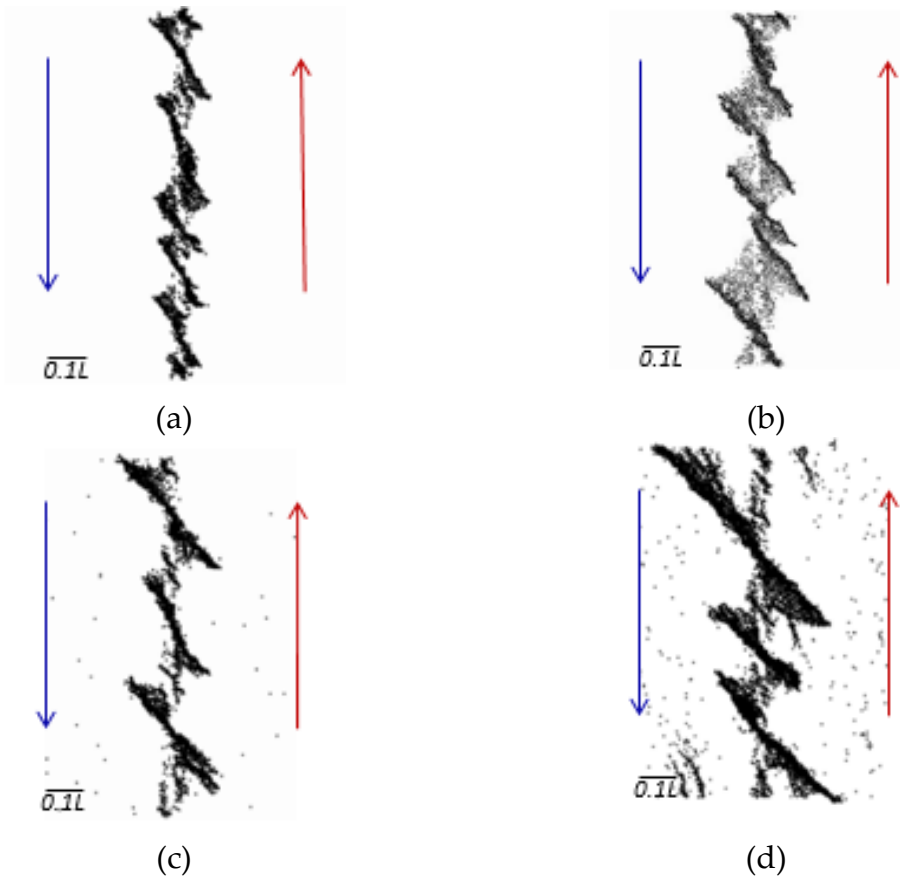

**Figure 5**. Top views of the numerical model used to simulate the propagation of en-echelon fractures in an intact elastic brittle layer with different thicknesses T subjected to Pure Strike-Slip; (**a**) PSS_0.66$T_{ref}$ (**b**) PSS_$T_{ref}$; (**c**) PSS_1.33$T_{ref}$, (**d**) PSS_2$T_{ref}$.

The evolution of the spacing *S* and the twist angle ω as functions of the thickness $T_{ref}$ are presented in **Table 2**. As the thickness T increases, ω increases from 31° towards a maximum limit value of 38°. In addition, the length *l* of the en-echelon fractures as well as the spacing *S* between the longer ones, increase significantly until a new step in the hierachization process occurs for a thicker layer. A consequence of the thickening is a lower level of interaction between the propagating fractures, easing the hierarchical process previously described (Chau and Wang, 2001; Goldstein and Osipenko, 2012). For the thickest layer tested (**Figure 5-d**), the central fracture stops growing as the two surrounding ones reach longer lengths: this behavior characterizes the bifurcation process, which has been predicted by an elastic fracture mechanics approach (Goldstein and Osipenko, 2012).

2.3.2 Influence of the deformation mode: transtensional and transpressional cases

To investigate the influence of the deformation mode on the propagation process, we modified the displacement imposed on the model boundaries to simulate transtensional and transpressional deformation modes. Convergence and divergence angles noted $θ_C$ and $θ_D$ respectively characterize the angular direction of the boundaries' displacements. An angle of zero degree generates a pure strike-slip deformation mode. For both transtensional and transpressional modes, angle of divergence and convergence *θ* of 10°, 20° and 30° were selected (**Figure 6**).

For the transtensional cases (**Figure 6-a,b,c,d**), as the divergence angle increases, the twist angle and the spacing between the en-echelon fractures decrease strongly (see **Table 2**). Because of the transparent top view, the connection at depth between neighboring fractures is visible. For the transpressive cases, i.e. as the convergence angle increases (**Figure 6-e,f,g,h**), the twist angle remains constant, as the spacing between the fractures increases slightly (**Table 2**). Note that the amount of cracks increases drastically as a result of to the increasing level of compressional stress.

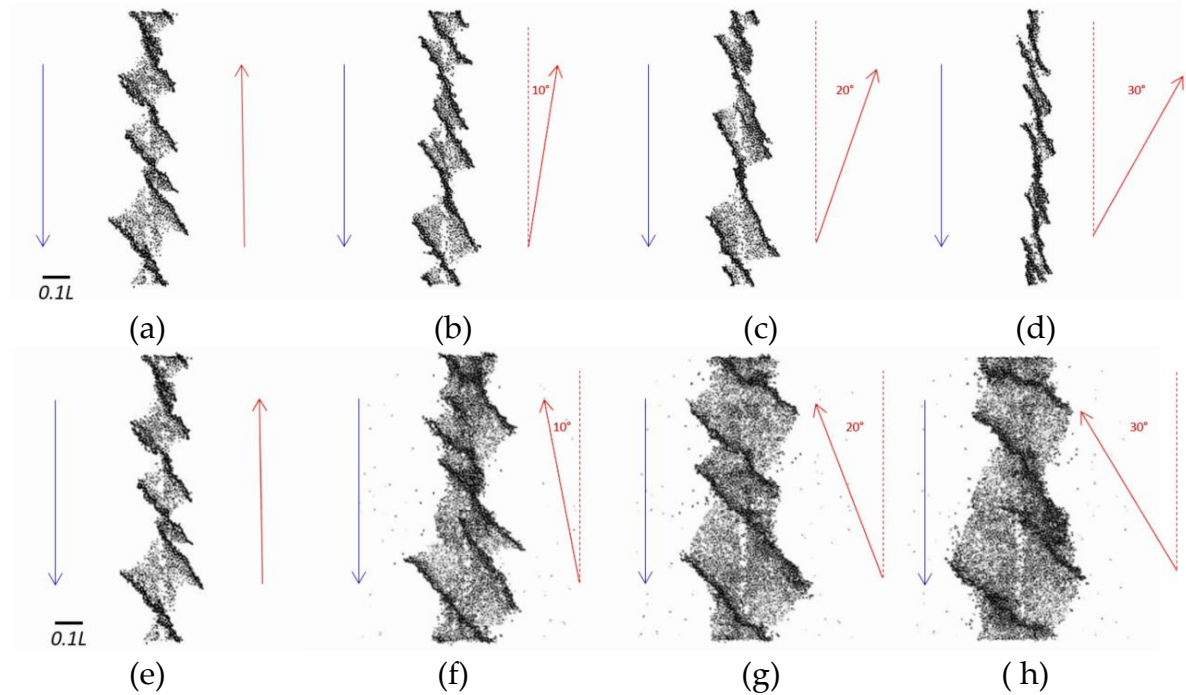

**Figure 6**. Top views of the numerical models used to simulate the propagation of en-echelon fractures in an intact elastic brittle layer of thickness $T_{ref}$ subjected to transtension (TT for TransTension) modes for different angles $θ_D$ of divergence with the associated test ID: (**a**) 0°, PSS_$T_{ref}$; (**b**) 10°, TT_10; (**c**) 20°, TT_20 and (**d**) 30°, TT_30. Bottom views, Transpression (TP) modes are controlled for different angles $θ_C$: (**e**) 0°, PSS_$T_{ref}$; (**f**) 10°, TP_10; (**g**) 20°, TP_20 and (**h**) 30°, TP_30.

**Table 2.** Numerical results. Standard deviations are plotted in Figure 7.

| Test ID | Thickness T | Average twist angle ω | Average Spacing s |
|---|---|---|---|
| PSS_0.66T$_{ref}$ | 0.67T$_{ref}$ | 31 | 1.76T$_{ref}$ |
| PSS_T$_{ref}$ | T$_{ref}$ | 38 | 3.2T$_{ref}$ |
| PSS_1.33T$_{ref}$ | 1.33T$_{ref}$ | 38 | 4.6T$_{ref}$ |
| PSS_2T$_{ref}$ | 2T$_{ref}$ | 40 | 6T$_{ref}$ |
| TT_10 | T$_{ref}$ | 24 | 1.66T$_{ref}$ |
| TT_20 | T$_{ref}$ | 28 | 2.13T$_{ref}$ |
| TT_30 | T$_{ref}$ | 30 | 2.53T$_{ref}$ |
| TP_10 | T$_{ref}$ | 41 | 2.53T$_{ref}$ |
| TP_20 | T$_{ref}$ | 42 | 2.66T$_{ref}$ |
| TP_30 | T$_{ref}$ | 43 | 2.93T$_{ref}$ |

*2.3 Comparison between the DEM model results with analytical formulations*

Based on fracture mechanics, several formulations linking the twist angle ω with the solicitation mode have been proposed in the literature (Pollard et al., 1982; Lazarus et al., 2008; Leblond et al., 2011; Pons et al., 2010). It is possible to compare the DEM results with these formulations, because the model simulates the behavior of a cohesive-frictional medium (see Scholtès and Donzé, 2013 for details). However, we have to express the degree of divergence given by the angle $θ_D$ in terms of the stress intensity ratio $\frac{K_{III}}{K_I}$. Note that a stress intensity factor K, is used to predict the stress state ("stress intensity") near the tip of a notch or a fracture caused by a remote load. $K_I$ applies to the stress intensity factor for mode I loading i.e. the opening mode and $K_{III}$ applies to the out of plane shearing mode III. Because the transtension mode in our model is controlled by the divergence angle $θ_D$, we can convert it into $\frac{K_{III}}{K_I}$ using the relationship proposed by Leblond & Frelat (2014),

$$\frac{1-\nu}{2}\cot(θ_D) = \frac{1}{\sqrt{2}}\frac{K_{III}}{K_I} \qquad (7)$$

Comparing the evolution of the twist angles obtained from two DEM simulations using different **ks/kn** with the analytical ones proposed by Leblond and Frelat (2014) and by Pollard et al. (1982) (**Figure 7**). It can be observed that the trends of the different results are similar.

## 3. Predicting the brittle layer thickness from the twist angle, spacing and length of the en-echelon pattern

As seen in the previous sections, the Riedels or en-echelon fractures can be characterized by the spacing *S* and the twist angle ω (**Figures 5 and 6**). Our objective was to set up a relationship between these two parameters according to the thickness T of the deformed layer. We first plot the different results of our numerical simulations in terms of fracture spacing over the thickness of the brittle layer ($\frac{S}{T}$) versus the twist angle ω (**Figure 8-a**). The main observed trend is that the value of the ratio ($\frac{S}{T}$) increases as the twist angle ω increases accordingly.

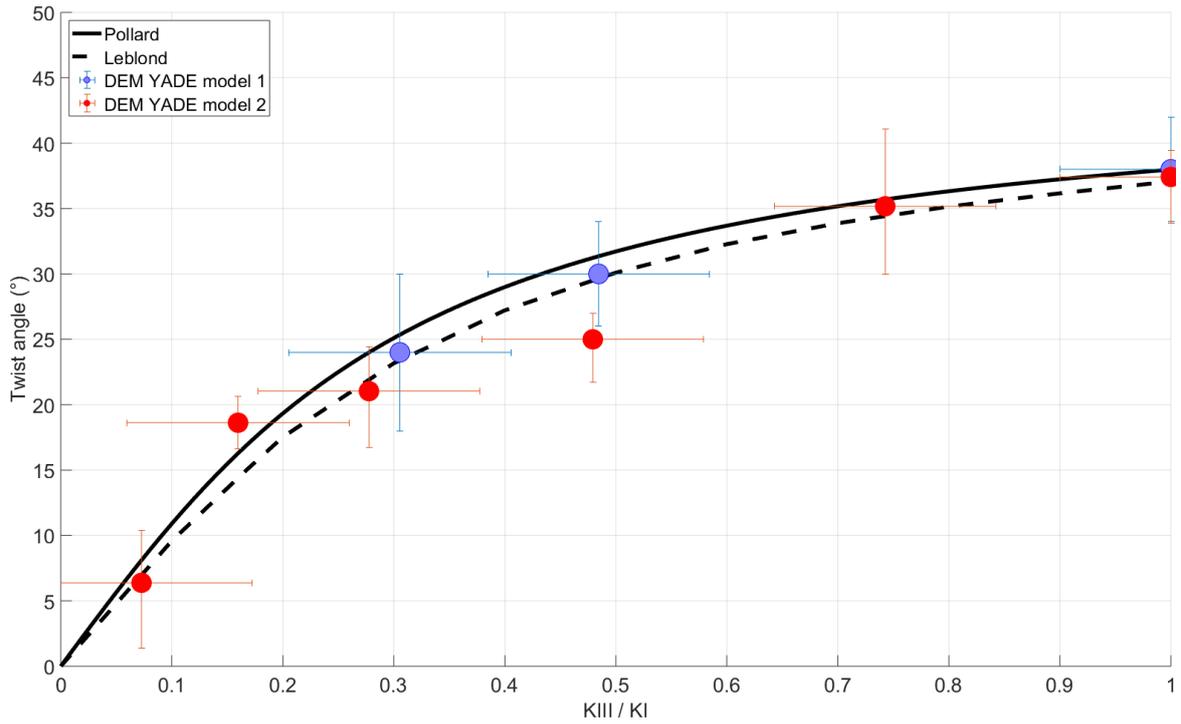

**Figure 7.** The evolution of the twist angle ω as a function of the stress intensity ratio $K_{III}/K_{I}$. The blue and red dots correspond to the numerical results obtained for two different models. The plain black line correspond to the formulation by Pollard et al. (1982), and the dash line to the formulation by Leblond and Frelat (2014).

Then, in addition to our data set, we back analyzed different small-scale experiments performed with clay (Tchalenko, 1970; Clifton et al., 2000; Schlische et al., 2002), sand (Naylor et al., 1986; Zuza, 2017; Lefevre et al., 2020), walnuts (Zuza, 2017) and gypsum (Goldstein and Osipenko, 2012) to get the evolution of the twist angle ω versus the fracture spacing over the thickness of the brittle layer ($\frac{S}{T}$). We plot the results in **Figure 8-b**. We deliberately mixed different types of media (clay, sand, walnuts and gypsum) in order to consider the en-echelon process whatever the nature of the medium is. From **Figure 8-b**, we can see that our numerical results fit the general trends of the observations. Globally, the ratio ($\frac{S}{T}$) evolves in a nonlinear manner with ω. For low values of ω, this ratio is also low. As ω increases, its evolutions express a sharp transition up to what appears to be an asymptotic limit, slightly higher than 3 for ($\frac{S}{T}$) (**Figure 8-b)**.

Expressing a mathematical formulation linking ($\frac{S}{T}$) with the twist angle ω based on a complete and independent set of physical parameters turned out to be out of reach in this study, despite our massive efforts. However, the main trend of the overall results can be fitted using a nonlinear least-squares solver based on a trigonometric function, which leads to the following expression:

$$\frac{s}{t} = 1.2 * \tan^{-1}\left(\left(\frac{5}{2}\omega - 1\right)\pi\right) + \frac{\pi}{2} \qquad (8)$$

The corresponding curve is plotted in **Figure 8-b**. One can note that the ratio ($\frac{S}{T}$) tends to be very small as the twist angle ω tends to zero. This trend has also been observed in several laboratory tests using Plexiglas: the en-echelon fractures exhibit a "lance" shape, i.e. the spacing between the fractures is several times smaller than their heights, which results in a "sawtooth" geometry (Lazarus et al., 2008; Pons and Karma, 2010; Cambonie et al., 2019).

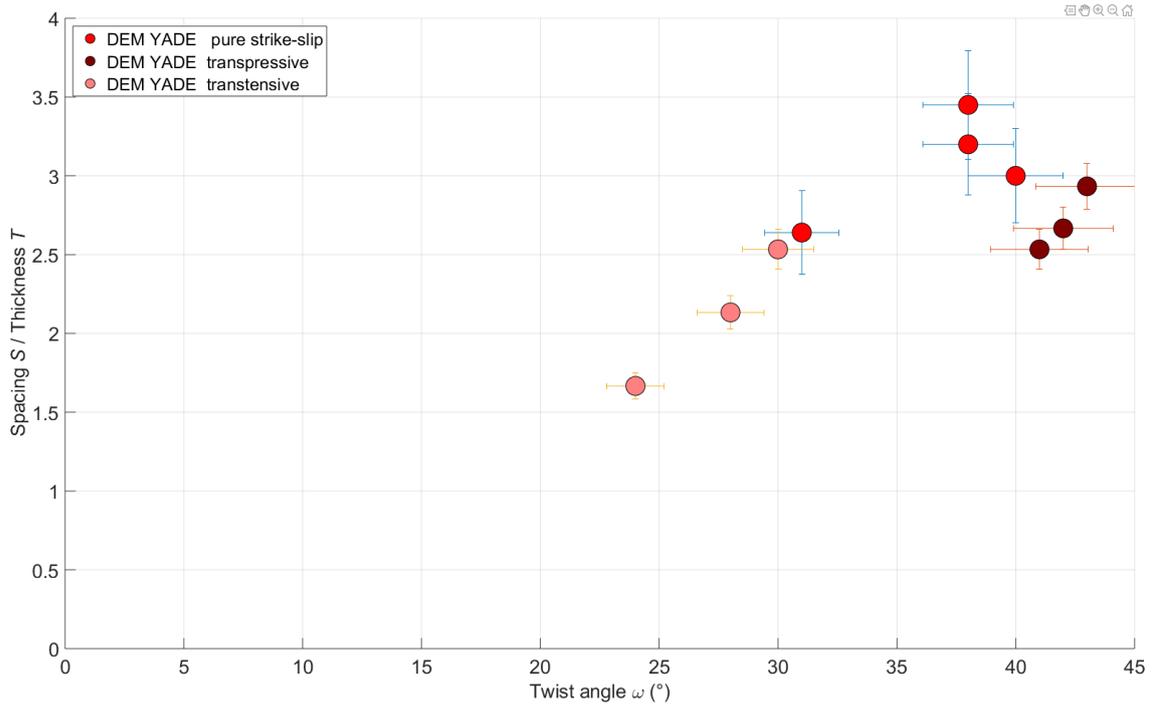

(a)

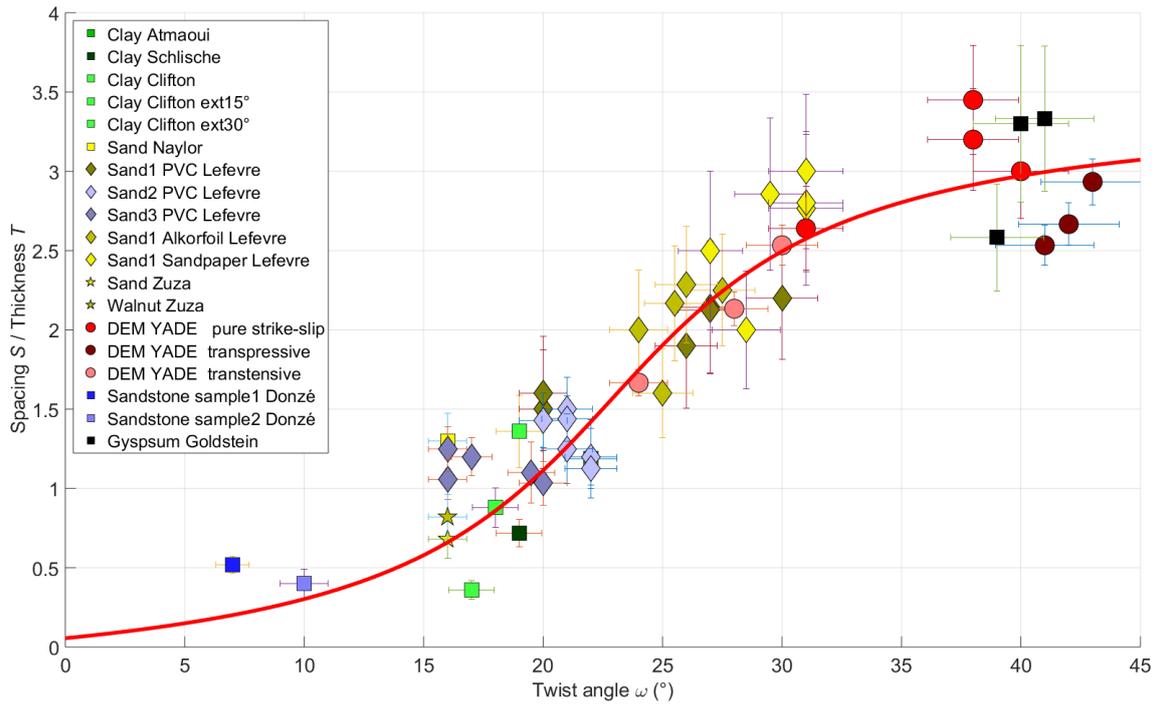

(b)

**Figure 8.** (a) The ratio between the spacing $S$ and the thickness $T$ is plotted versus the twist angle for the results obtained with the discrete element model presented in Figures 5 and 6. (b). Added to the discrete element model, results obtained from different small-scale experiments performed with clay (Tchalenko, 1970; Clifton et al., 2000; Schlische et al., 2002), sand (Naylor et al., 1986; Zuza, 2017; Lefevre et al., 2020), walnuts (Zuza, 2017) and gypsum (Goldstein and Osipenko, 2012) are also plotted. The red curve corresponds to the trigonometric function given in Equation 8 fitting all the data set.

## 4. Applying the relationship linking ($\frac{S}{T}$) to the twist angle ω to the tectonic faults

Continental strike-slip faults can cut through the entire lithosphere in which only the upper layer exhibits a brittle behavior, i.e. the first ~15–20 km from the surface where most of the seismogenic activity occurs. These large continental faults accommodate plate-tectonic movements (Tapponnier et al., 1982), and they correspond to large-scale fractures propagating in a brittle layer from shear deformation occurring at a certain depth. During the genesis stage of the fault, the fracture grows upward toward the surface, forming en-echelon patterns or Riedels, oriented at an angle ω to the regional basal shear displacement. As the fault is developing, the main sliding zone localizes forming P shears (**Figure 2-b,c,d,e**) which cross and connect the Riedel system as the en-echelon patterns formed by the Riedels become inactive. If the location of the original P-shears (**Figure 2-c**) delineates the segments along the main fault, the fault segments observed today would be the reminiscence of the initial fault pattern formed during its genesis (Klinger, 2010; Lefevre et al., 2020). In the next section, we discuss what we can expect in terms of seismic depth prediction from the en-echelon patterns at surface along some major well-developed faults.

*4.1. En-echelon pattern related to co-seismic deformation: Darfield (Canterbury, New Zealand) earthquake 2010 along the Greendale fault case*

The Mw 7.1 Darfield (Canterbury) earthquake of 4 September 2010 produced ground-surface trace of the previously unrecognized Greendale Fault (Quigley et al., 2010) (**Figure 9-a**). From fault slip models of the 2010 Canterbury earthquake zone (**Figure 9-b**), Beavan et al., (2012) suggested that the depth of the maximum slip zone ranges in depth somewhere between 2 000 m to 5 000 m for the Western area and from 4 000 m to 9 000 m for the Eastern part. Another model based on strong ground motions, assessed the maximum slip zone depth between 3 000 m to 12 000 m for the Western part between 4 000 m and 14 000 m for the Eastern one (Irikura et al., 2019). It has been assumed that this moderately shallow earthquake was produced by the reactivation of a blind parent fault covered by intact layers of sediments (Arthur and Lawton, 2013). This case is of major interest in our present study because en-echelon series of East-West striking, left-stepping fracture traces extending around 30 km were observed (**Figure 9-b**).

Vertical aerial photographs were acquired over 20 km of the Greendale Fault six days after the event, providing an accurate localization of the rupture patterns at the surface. To determine the value of the twist angle, we have considered the largest step-overs north of Burnham, near Rolleston and we drew the axis crossing the Riedels along the incipient P shear (**Figure 9-b**).

On the Eastern zone, the twist angles ω measured for the two main Riedels vary from 15° to 18°. The spacing measured between these Riedels is about 5330 m (**Figure 9-b**). Using Equation 8, we estimate that the corresponding depth of a "potential" parent fault below the broken brittle layer would be around 6 250 m – 9 200 m. This range of values is between the ones predicted by both Beavan et al., (2012) and Irikura et al. (2019) (**Figure 10-a,b**). It means that the lower boundary of the fractured layer could be inside the undifferentiated Mesozoic basement (Arthur and Lawton, 2013) that has probably not moved for a very long period (Beavan et al., 2012), so that it ruptures like an intact medium.

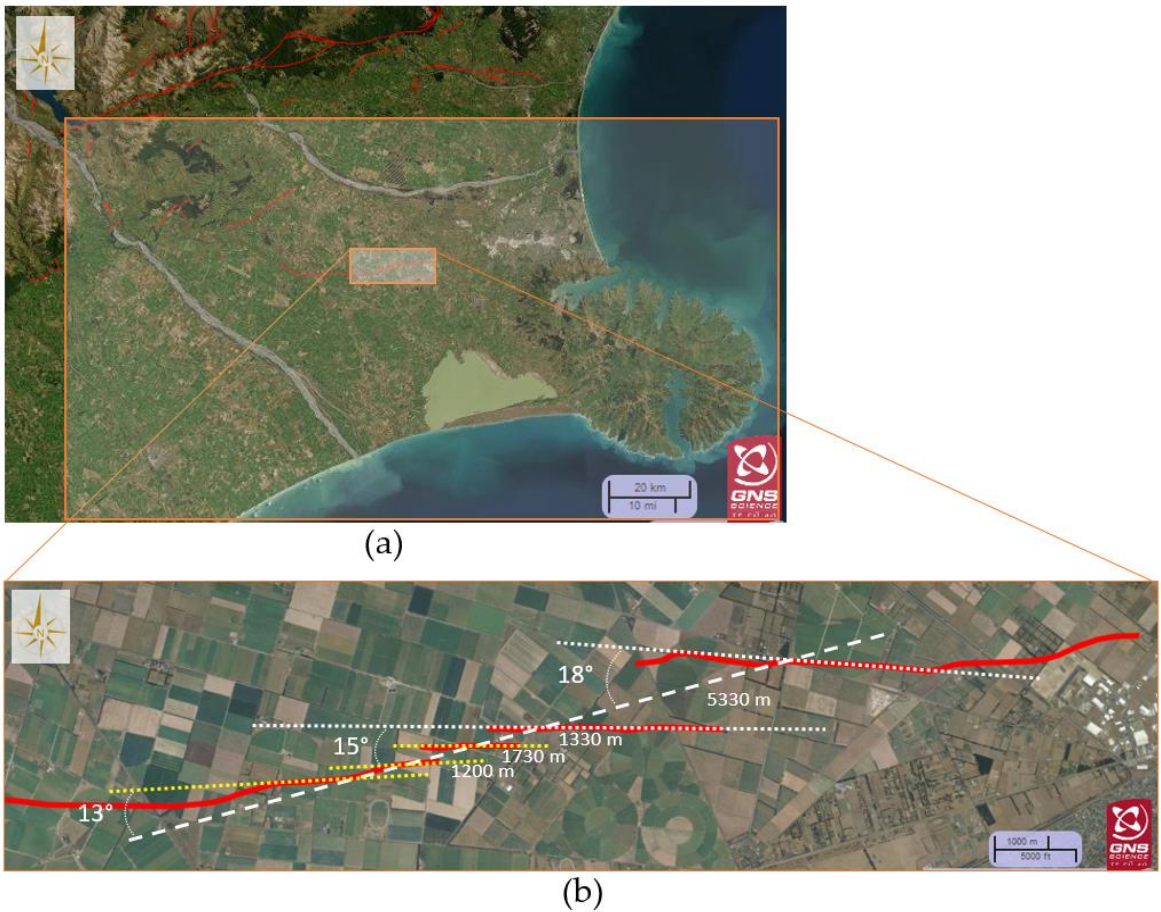

**Figure 9.** (a)Surface traces of onshore active faults in the Christchurch area with the Eastern section of the Greendale fault (source: GNS science) (b) Spacing distances between the longest en-echelon Riedels identified on the West of the hamlet of Greendale (the distances are measured along the direction of the estimated P shear).

On the Western zone, the Riedels, highlighted with a yellow dashed line in **Figure 9-b**, are less spaced from each other. With an average twist angle $\omega$ of 14° and spacings varying from 1 200 m of 1730 m, the corresponding thickness of the fractured layer would be about 2 360 m to 3 400 m. These shallower values correspond to the lower depth of the maximum slips suggested at 2000 m by Beavan et al. (2012) and at 3000 m by Irikura et al. (2019).

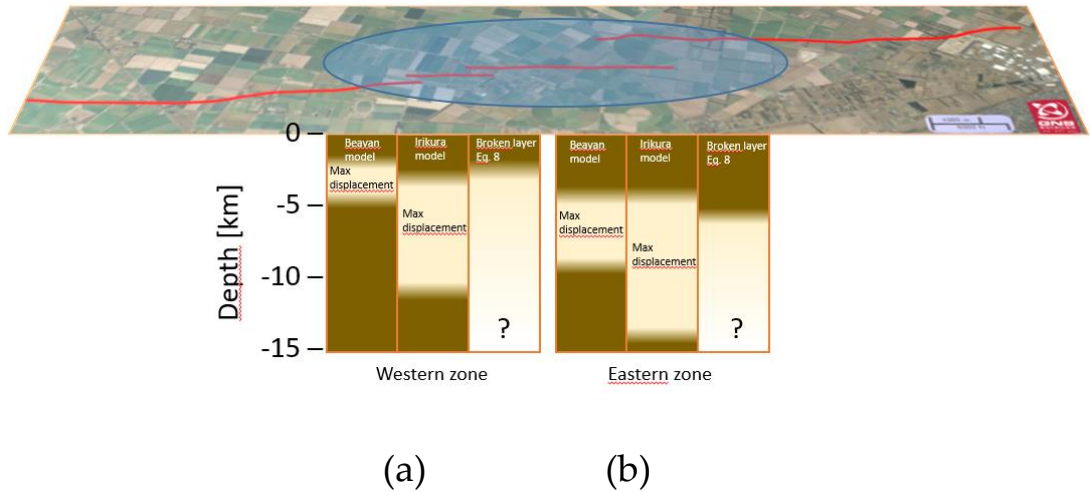

(a)        (b)

**Figure 10.** Western (a) and Eastern (b) zones of the Greendale fault represented in **Figure 9**. For (a) and (b) respectively, the first column represents the maximum displacement zone according to Beavan et al. (2012) and Irikura et al. (2019) in the second column. In the third column, the dark zone represents the fractured brittle layer above the parent fault at depth (clear zone). The question mark illustrates the fact that Equation 8 can only provide the thickness of the fractured layer above the parent fault at depth.

*4.2. Well-developped strike-slip faults exhibiting segments*

As mentioned above, when strike-slip faults develops, the initial Riedels become inactive and the deformation localizes along the P Shear (**Figure 2-b,c,d,e**), forming relay zones. Because the P Shear zone can cross or follow the initial Riedels, we can eventually assume that the change of direction of the fault segments can be related to the initial twist angles. Based on this assumption, Equation 8 can then be used to estimate the thickness of the brittle crust which has been fractured by some major strike-slip faults. Here we consider the following faults for which the lengths of segments have been measured (Klinger, 2010): Johnson and Homestead Valleys including Landers (United-States), Hector Mines (United-States), Superstition Hills (United-States), Zirkhu (Iran), Luzon (Philippine) and Gobi-Altay (Mongolia) (faults listed in **Table 3**). We have selected these faults because of the high quality maps of their surface traces, but also because they are recent enough for us to get information about their seismogenic thicknesses (**Table 3**). Based on RMS - misfit trend considerations, the longer segments used to describe the fault segmentation are considered as input data in Equation 8. In addition, we have measured the values of the average azimuth change between consecutive segments for each of these faults (**Table 3**).

Using Equation 8, we calculated the possible thicknesses of the brittle crust crossed by these faults and compared them to their seismogenic thicknesses reported in the literature (**Table 3**). Because of the high variability of the data set and the possible biases induced by our measuring technique, we obviously remain cautious in our conclusions.

Nonetheless, we observe that the thicknesses of the brittle rupture predicted by Equation 8 are close to the ones corresponding to the seismogenic thicknesses for Johnson and Homestead Valleys, Hector Mine and Superstition Hills cases.

Table 3. List of faults considered in this study.

| Faults | Average length of segments (km) with standard deviation | Mean angle (°) ± 1° | Predicted thickness of the brittle breaking (eq. 8) | Seismogenic depth range (km) |
|---|---|---|---|---|
| Johnson and Homestead Valleys (United States) | 19 ± 0.9 [a] | 22° | 12.7-14.6 | 15[b] |
| Hector Mine (United States) | 10.25 ± 2.71 [a] | 20° | 6.77-11.63 | 10-23[b] |
| Superstition Hills (United States) | 9.3 ± 0.7 [a] | 20° | 7.7-8.96 | 9[b,c] |
| Zirkhu (Iran) | 20 ± 6 [a] | 15° | 24.17-44.9 | 18-20[d] |
| Luzon (Philippine) | 19 ± 4.7[a] | 12° | 36.41-60.35 | 15-25[e] |
| Gobi-Altay (Mongolia) | 16.24 ± 5.66[a] | 17° | 14-29 | 20[f] |

[a](Klinger, 2010), [b](Nazareth and Hauksson, 2004), [c](Magistrale et al., 1989),[d](Marchandon et al., 2018),[e](Velasco, 1996), [f](Kutz et al., 2018)

Regarding the Zirkhu, Luzon or Gobi-Altay cases, it seems that the measured low values of the average azimuth changes between consecutive segments induce an overestimation of the thickness of the brittle crust. Do these azimuth angles remain constant in time? Not always apparently and they probably decrease in time as the fault tends to present a large lateral displacement. Tracking the inactive branches, which could correspond, to the primary Riedels along the fault could represent a key option to improve the representativity of the selected twist angle values. However, these faults are generally not visible anymore.

**5. Possible application of the predicting law to strike-slip faults on Mars**

In the absence of plate tectonics and significant erosion process during the last four billion years, the landscape of Mars is still dominated by two structures dating back to the epoch of large-scale impacts, namely the north-south dichotomy and the Tharsis bulge (Beuthe et al., 2012; Bouley et al. 2020). Large volcanic plains and shields characterize the Tharsis region. In its eastern part, Valles Marineris radially cuts the Tharsis bulge. South-West of Tharsis, stratigraphic and tectonic evidence suggests that Memnonia-Sirenum region also formed concurrently with the early formation of Tharsis when the dynamo was still active (Anderson et al., 2019). Detailed geologic investigations have revealed distinct basins and ranges in the Memnonia-Sirenum region.

In both, Memnonia-Sirenum region and Valles Marineris, strike-slip fault systems have been identified (Yin, 2012; Andrews-Hanna et al., 2008). In order to estimate the thickness of the brittle layer crossed by these faults, we selected a strike-slip fault system exhibiting en-echelon structures.

*5.1. Strike-slip fault in the Memnonia region*

Andrews-Hanna et al. (2008) have reported evidences of strike-slip faults in the Southwest region of Tharsis. Most of the potential strike-slip faults have orientations relative to the principal stress directions which occured during the Noachian (Anderson et al., 2001) i.e. when the bulk of Tharsis construction occurred. We have selected one of these, located North-West of the Burton crater (Andrews-Hanna et al., 2008) (**Figure 11-a**) which exhibits an en-echelon pattern.

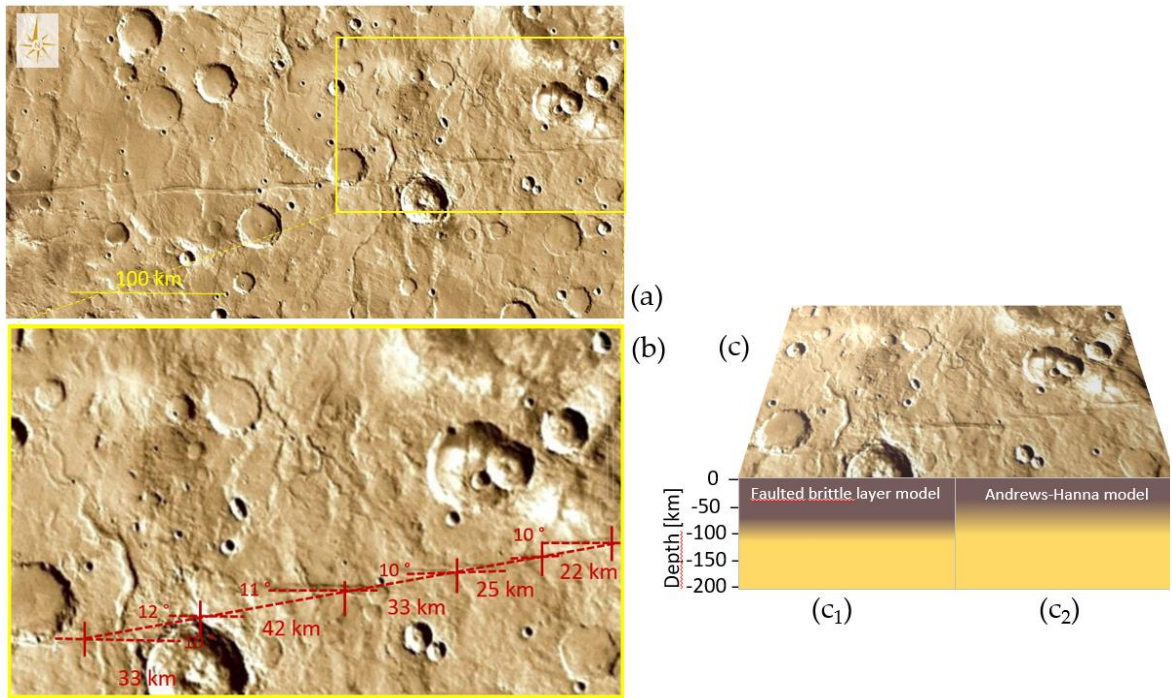

**Figure 11. (a)** Potential transtensional strike-slip fault along a Graben zone in the Memnonia region to the southwest of Tharsis (13°52'29.25"S, 160°39'18.19"O), West of Burton crater (Image NASA/USGS/DLR/FU Berlin). **(b)** The spacing distance between the en-echelon segments oriented along the dash line with the associated twist angles). **(c)** In dark brown, the estimated brittle crust thickness of 72-101 km predicted by Equation 8 ($C_1$). Also in dark brown, the crustal thickness 35 and 100 km estimated by Andrews-Hanna et al. (2008) based on thermal evolution models ($C_2$). Note that the gradient between two colors indicates the level of uncertainty.

The very low twist angles between the potential Riedels visible at the surface **(Figure 11-b)** suggest that the associated mechanism should be mostly transtensive shearing, which seems compatible with the Western part of the fault zone where the segments become aligned, composing an extensive fault system. This possibly forms the edge of a Graben. The Riedels align along a main line, with a regular spacing ranging from 22 km to 42 km. With twist angle values of about 10°-12°, the corresponding thickness obtained from Equation 8, is about 72-101 km (**Figure 11-$C_1$**). Using the range of contractional strains predicted from thermal evolution models, Andrews-Hanna et al. (2008) constrained the brittle crust thickness at the time of Tharsis rise, between 35 and 100 km (**Figure 11 $C_2$**). Other studies suggest that the crustal thickness for Memnonia region is about 60-80 km based on present-day heat flow model (Parro et al., 2017) and 50-60 km from gravity and topographic models (Tenzer et al., 2015).

## 6. Extrapolating the predicting law to en-echelon fault systems on Icy moons

*6.1. Enceladus*

At the South Pole of Enceladus, four sub-parallel, linear depressions compose the Tiger stripes structure. The four stripes are made of a series of sub-parallel, linear depressions flanked on each side by low ridges (Yin et al., 2016). Their correlation with internal heat and a large, water vapor plume suggests that the Tiger stripes might be the result of faulting in Enceladus' lithosphere. To measure spacing between the stripes and to eventually estimate a twist angle, we need to identify the direction of the maximum shear deformation loading of Enceladus's South polar region. The difficulty to get this value is that this shear loading is highly dependent on the periodic tidal deformation of the ice shell. It turns out that the shear deformation is preceded by an extension phase and followed by a compressive one. Assuming that the shear deformation is of the first order, we estimated this direction from the mechanical model proposed by Souček et al. (2016) (**Figure 12-a**). Doing so, we could measure a twist angle of about 55° and a spacing between the stripes of about 35 to 45 km. This high value of the twist angle could be due to the complex loading sequences resulting from the Tidal deformations. This means that applying Equation 8 would provide a minimum value for the thickness of the outer brittle shell only (Rhoden et al., 2019). We find that the thickness of the possible brittle layer associated to this geometry is about 11-14 km (**Figure 12-b$_1$**). This predicted value of the thickness could be compatible with several scenarios proposed by authors using different modeling approaches, including topography or libration amplitude (Yin et al., 2016) (**Figures 12-b$_2$,b$_3$**) also based on a mechanical approach (Van Hoolst et al., 2016) or based on physical libration amplitude (Čadek et al., 2016; Rhoden et al., 2019).

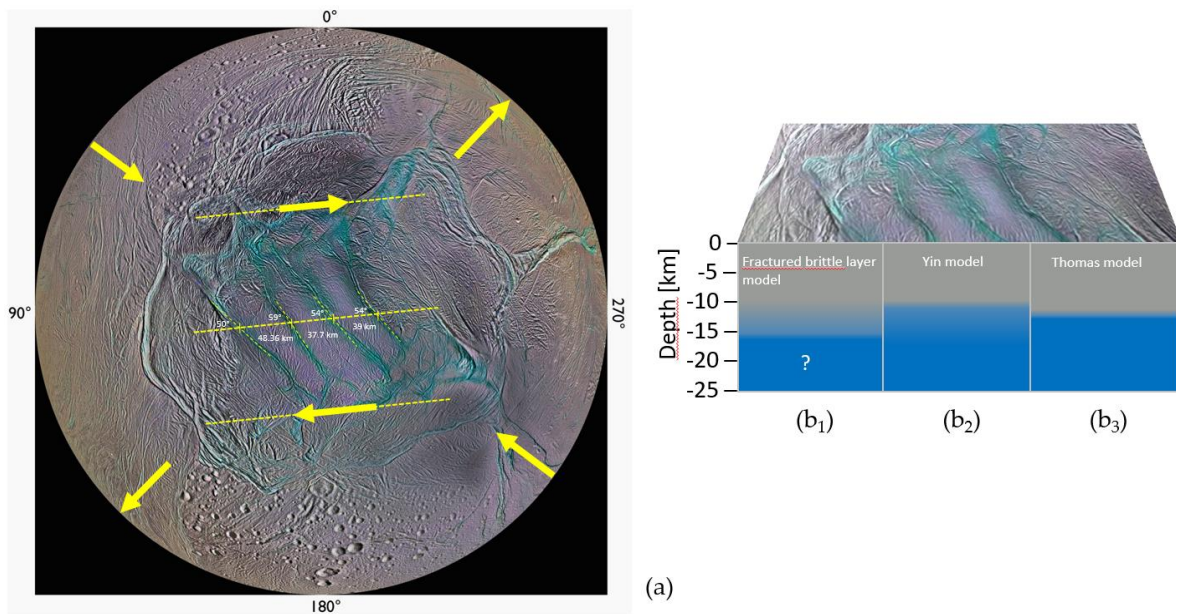

**Figure 12.** Simplified tectonic map of the South Polar Terrain based on the analysis of images obtained by Cassini orbiter's Imaging Science Subsystem (NASA/JPL-Caltech/SSI/Lunar and Planetary Institute, Paul Schenk (LPI, Houston). Tiger-stripe fractures in the South Polar Terrain of Saturn's moon Enceladus (a). In gray, the minimum thickness of the brittle layer obtained from Equation 8, 11 km-14 km (b$_1$), vs. the 10 km predicted by Yin et al.(2016) (b$_2$), and 13 km predicted by Thomas et al. (2016) (b$_3$). Note that the gradient between two colors indicates the level of uncertainty.

*6.2. Ganymede*

The high-resolution Galileo data set of Ganymede's complex surface provides strong and ubiquitous evidence of strike-slip motion and several en-echelon structures (Cameron et al., 2018) (**Figure 13, top**). According to these authors, it was proposed that tidal stress models of combined diurnal and nonsynchronous rotation stress readily promote brittle behavior of the external crust layer with a corresponding thickness of 2 km (**Figures 13-b,d**). We considered an en-echelon structure identified in the light subdued material found in the Transitional terrain ( 173°E, 32°N), with a twist angle of 22° and a spacing of 3 to 4 km (**Figure 13-a**). According to Equation 8, the associated thickness for a brittle behavior is about 2.1 km-2.8 km (**Figure 13-a**). Within the dark lineated terrain of Anshar Sulcus (162°E, 18°N), the identified en-echelon structure exhibits a twist angle of 34° and a spacing of about 5 to 6 km. The corresponding thickness of the brittle layer is about 1.8 km-2.2 km (**Figure 13-c**).

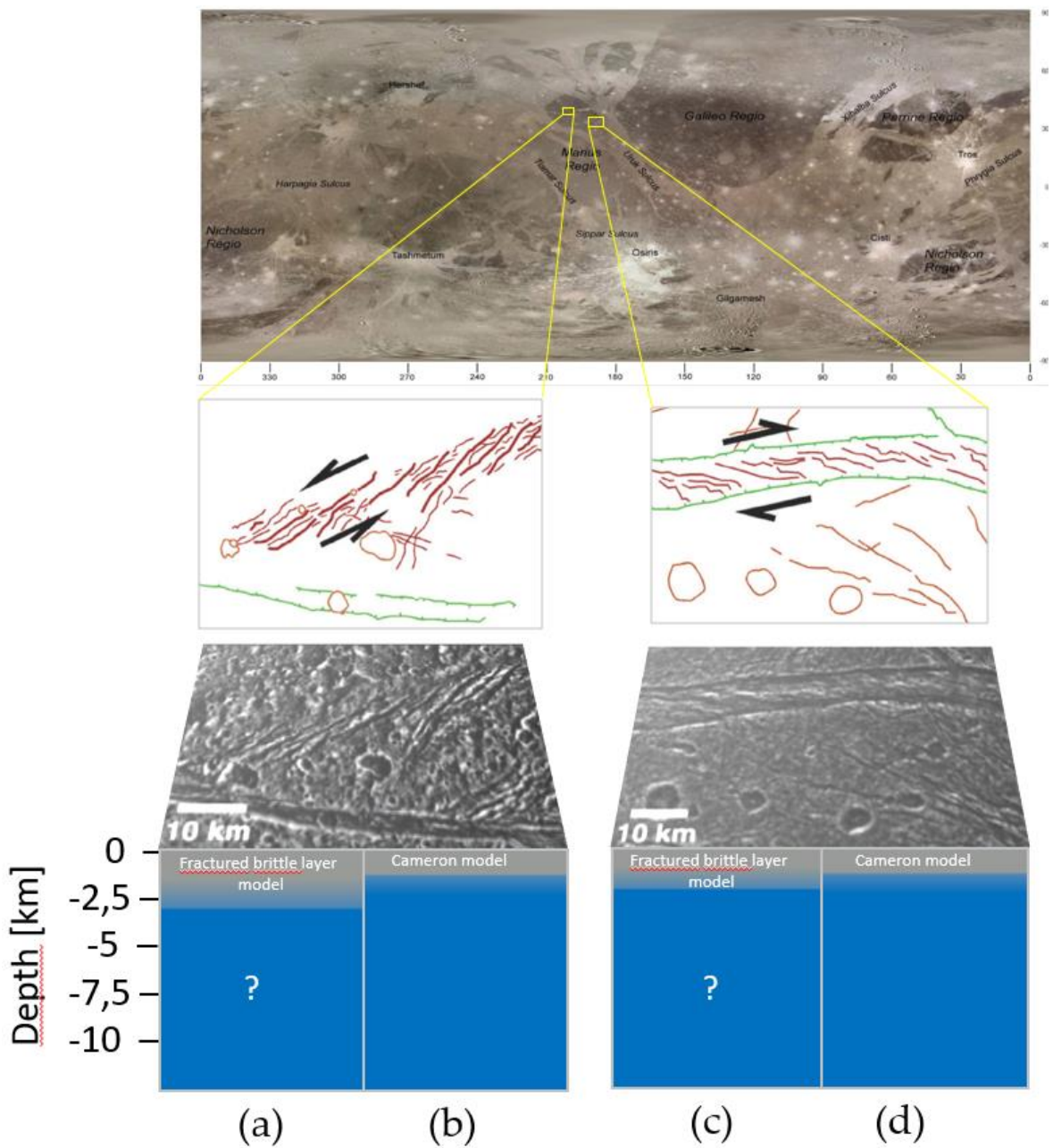

**Figure 13.** On Top, Ganymede global mosaic in Mercator projection (NASA/ESA/JPL/SSI/Cassini Imaging Team). Bottom left: En-echelon structures within the dark lineated terrain of Anshar Sulcus (162°E, 18°N); In Gray the thickness predicted by Equation 8 (a) vs. the prediction given by Cameron et al., (69-2019) (b). Bottom right, en-echelon structures in the light subdued material found within Transitional Terrain (173°E, 32°N). In Gray the thickness predicted by Equation 8 (c) vs. the prediction given by Cameron et al., (2019) (d). Arrows indicate inferred shear sense assuming a transtensional origin, where bold red lines are features of interests, thin red lines are craters and features outside of the relevant zone, and green lines are scarps (ticks point downslope) (from Cameron et al., (2018), Deremer et al., (2003)). Note that the gradient between two colors indicates the level of uncertainty.

*6.3. Europa*

Europa's surface is tectonically and morphologically complex. Europa ridges, bands, ridged bands, double ridges, complex ridges, and troughs are collectively referred to as lineaments (Michalski, 2002). Some lineaments are fault zones and exhibit sets of en-echelon ridge and trough structures. We studied three sets of these en-echelon structures presented by Michalski, (2002) page 136, Figure 1) (**Figure 14 left**). Twist angles range from 21° to 25° while spacings range from 2.76 to 4.6 km. The resulting thicknesses calculated with Equation 8, ranged from 2.2 km to 3.6 km (**Figure 14 right**). Several models based on mechanical methods provide similar ice shell thickness (Billings and Kattenhorn, 2005; Chuang et al., 2001), but they are challenged by other predictions (for example, 3 to 30 km proposed by Kalousova, (2017), which could suggest that we are in presence of paleo-structures or there is a high variability of the ice shell thickness of Europa.

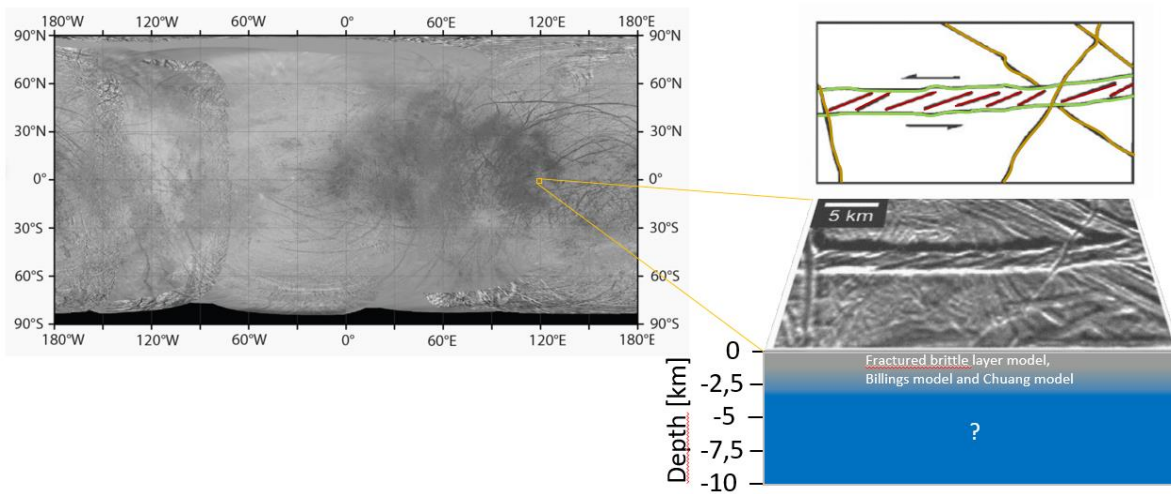

**Figure 14.** On the Left, Controlled Photomosaic Map of Europa, Je 15M CMN. (U.S. Geological Survey Geologic Investigations Series I–2757). On the Right: En-echelon ridges identified on mosaic of SSI images s0449961800 and s0449961865, 246 m/pixel, located at 22°N,223°W, (Michalski, 2002). In Gray the possible thickness of the ice shell predicted by Equation 8 in agreement with other models based on mechanical methods thickness (Billings and Kattenhorn, 2005; Chuang et al., 2001).

**7. Concluding remarks**

Using a numerical model based on the discrete element method, we have simulated the en-echelon fractures formation. The results are in good agreement with the hierarchical process proposed by Goldstein and Osipenko (2012) which can explain the relationship between the spacing between the Riedels and the thickness of the fractured brittle medium. Moreover, the DEM model reproduced the dependency of the twist angle value on the mode of deformation for transtensive cases, similar to the ones obtained by fracture mechanics formulations (Pollard et al. 1982; or by variational methods (Leblond and Frelat, 2014).

We have thus compared our numerical results to large amount of results obtained through experimental models, using clay, sand, walnuts and gypsum materials. A relationship between the the twist angle ω of the Riedels and the ratio between their spacing and the thickness of the brittle layer has been set up (Equation 8) based on a nonlinear least-squares fitting. This relationship has been used to assess the thickness of the fractured brittle crust crossed by tectonic faults.

For well-developed tectonic faults, we assumed that the length of the segments visible on well-developed faults could be related to the P shears connecting and/or crossing the initial Riedels. The orientation of these segments might be partly controlled by the persistence of initial twist angle of the Riedels but as the displacement along the fault increases, they no longer appear to be related.

The application of the proposed relationship (Equation 8) to Mars or the icy moons is highly speculative and we face here a major challenge because we do not know the thicknesses of the brittle crusts. We also ignore whether the thicknesses of the ice shelfs of the icy moons have been constant through time and if, different layers of different mechanical properties are present, e.g. brittle-ductile boundary transition due to the presence of potential radiative heat flows. Questions also arise for the structure of Mars, and we are looking forward to the first tomography of its interior.

**Funding**

This research was funded by Agence Nationale de la Recherche, "Geometry of Strike-Slip faults through Multiple Earthquake Cycles – GeoSMEC" project, grant number ANR-12-BS06-0016.

**Acknowledgements**

Grateful thanks to the Editor and the two reviewers who spared no effort to help us improve the quality of the article. The first author would like to thank Sophie-Adélaïde Magnier for helpful discussions, as well as Hervé Dondey and Bernard Schmitt for fruitful interactions during a field work in High Atlas, Morocco.